\documentclass[referee]{raa_rb}

\usepackage{graphicx,times}
\usepackage{natbib}
\usepackage{amssymb,amsmath}
\bibpunct{(}{)}{;}{a}{}{,}

\usepackage[pagebackref=true]{hyperref}

\usepackage{tabularray}[=v2021]
\usepackage{multirow}

\begin{document}
\title{Observation-Based Iterative Map for Solar Cycles. II. The Gnevyshev-Ohl Rule and its Generation Mechanism}

   \volnopage{Vol.0 (202x) No.0, 000--000}      
   \setcounter{page}{1}          

   \author{Zi-Fan Wang 
      \inst{1,2}
   \and Jie Jiang
      \inst{3,4,*}\footnotetext{$*$Corresponding Author}
   \and Jing-Xiu Wang
      \inst{1,2}
    }

         \institute{State Key Laboratory of Solar Activity and Space Weather, National Astronomical Observatories, Chinese Academy of Sciences, Beijing 100101, China\\
        \and
             School of Astronomy and Space Science, University of Chinese Academy of Sciences, Beijing, China\\
        \and
             School of Space and Earth Sciences, Beihang University, Beijing, China; {\it jiejiang@buaa.edu.cn}\\
        \and
             Key Laboratory of Space Environment Monitoring and Information Processing of MIIT, Beijing, China\\
\vs\no
   {\small Received 202x month day; accepted 202x month day}}




\abstract{The Gnevyshev-Ohl (G-O) rule, or even-odd effect, is an important observational phenomenon in solar cycles, originally suggesting that even-numbered cycles are typically followed by stronger odd-numbered ones. However, subsequent studies have reported varied forms and often conflicting manifestations of this rule, along with diverse interpretations of its physical origin. Using an observation-based iterative map, we investigate these different forms of the G-O rule and propose a more general underlying rule: statistically, a given solar cycle is more likely to be followed by a stronger one, regardless of even-odd numbering. This transition asymmetry arises from the map's inherent asymmetry relative to the diagonal. Over timescales comparable to historical observations, both the G-O rule and its reversal can arise randomly, without a consistent preference. The short-term behavior of the rule is sensitive to the initial cycle, the selected time interval, and the parameters of the recursion function. These findings reconcile previously conflicting reports and point to a general generation mechanism: G-O-like behavior arises naturally from nonlinear stochastic dynamics. While different recursion parameters may lead to varying short-term patterns and statistical strengths, the emergence of G-O-like features appears to be a generic property of such systems.
}

   \authorrunning{Z.-F. Wang et al. }            
   \titlerunning{Genveyshev-Ohl rule and mechanism} 
\maketitle
\section{Introduction}\label{sec:intro}

Since the discovery of the magnetism in sunspots and its alternating sign across adjacent 11-year solar activity cycles by \cite{Hale1919}, the 22-year Hale cycle has been one of the most important topic in solar physics.  A critical observation phenomenon associated with the Hale cycle is the Gnevyshev-Ohl (G-O) rule, also known as the even-odd effect. As originally identified by \cite{1948G_O},  when solar cycles are paired by index, the following odd cycle is stronger than the previous even cycle, a pattern observed for all cycle pairs starting from the 18th century, with the exception of cycles 4 and 5. Hereafter we refer to this as the cycle strength definition of the G-O rule. \cite{1948G_O} also found that the correlation between even cycles and their following odd ones is significantly higher than that between even cycles and their preceding odd cycles, which we will hereafter referred to as the correlation definition  of the G-O rule.  Besides the original definitions by \cite{1948G_O}, cycle alternation, which describes that the cycle amplitudes tend to form a strong-weak alternating pattern \citep{2007ApJ...658..657C,2024Univ...10..364P}, is also a phenomenon highly related to the G-O rule, and is sometimes considered a definition of the G-O rule as well.  The G-O rule seemingly shows that the Hale cycle is a fundamental component of the evolution of solar cycles, and raise the important question about its physical origin.

With advancements in observational data and analytical methods, interpretations of the Gnevyshev-Ohl (G-O) rule have diverged in both methodology and results. A factor contributing to these discrepancies is the different representations of cycle strength. It can be represented either by the total sunspot number, as originally defined by \cite{1948G_O}, or by the maximum sunspot number, i.e., the cycle amplitude \citep[e.g.][]{2005LRSP....2....2C,2012SoPh..281..827J}.  The latter representation leads to more violations of the G-O rule \citep{2015LRSP...12....4H}, and reduces the statistical significance of the correlation definition \citep{2024ARep...68...89N}. Another point of divergence concerns whether solar cycles should be paired starting with an even or odd cycle. \cite{Turner1925} proposed that cycles should be paired starting with the stronger odd cycle, whereas \cite{2015Ge&Ae..55..902Z} argued that combination of solar cycles in pairs according to their numbers lacks a physical basis. The temporal range over which the G-O rule holds is also debated. \cite{2001SoPh..198...51M} showed that the G-O rule is in reversed phase between the Maunder and Dalton minima in the cycle strength definition.  \cite{2001A&A...370L..31U} and \cite{2009ApJ...700L.154U} suggested that a lost cycle in late 18th century caused the reversal of the G-O rule, while \cite{2013ApJ...772L..30T} suggests that the G-O rule may periodically change between larger even or larger odd cycles. On longer time scales, the validity of the G-O rule remains uncertain. \cite{2023IAUS..372...70S} found no strong statistical evidence for the G-O rule in a millennium-long solar cycle series reconstructed from cosmogenic radioisotopes, possibly due to the significant uncertainty in the reconstructed cycles as explained by the authors.  Thus, a comprehensive and definitive description of the G-O rule’s form and validity remains elusive.

The physical origin of the G-O rule is closely related to the origin of the solar magnetic field.  The formation and evolution of solar large-scale magnetic field are explained by solar global dynamo theories, in which the large-scale field arises from mutually generating poloidal and toroidal fields \citep{1955ApJ...122..293P, Charbonneau2020}.  Early explanations for the origin of the G-O rule often attributed it to the interaction between a non-alternating fossil field and the alternating dynamo field \citep{2001SoPh..198...51M}.  However, \cite{2005ApJ...619..613C} argued that as there are possibly reversals of the G-O rule, fossil field explanation is not favored.  Instead, they proposed that the G-O rule can be present in a nonlinear and stochastic dynamo process, independent of fossil fields.  In framework of the solar Babcock-Leighton (B-L) type dynamo models \citep{Babcock1961,Leighton1969}, the generation of toroidal field from poloidal field represented by the polar field at cycle minimum is generally linear \citep{Schatten1978,1979stp.....2..258O,2007MNRAS.381.1527J}, while the generation of poloidal field from toroidal field is intrinsically nonlinear and also stochastic \citep{Jiang2013,Jiang2020, Karak2020}.  \cite{2005ApJ...619..613C} demonstrated that the nonlinear and stochastic mechanisms introduced in a B-L dynamo can generate a G-O rule consistent with the cycle strength definition.  \cite{2005ApJ...619..613C} further shows how cycle alternation can happen from various ranges of nonlinearity, perturbed by stochasticity. \cite{2013ApJ...772L..30T} conducted simulations with an $\alpha\Omega$ dynamo including nonlinearity and stochastic effects, and found that it also generates properties consistent with the G-O rule. Despite these advancements, the varying  interpretations of the G-O rule’s form cause ongoing debate on its origin.  

When analyzing the dynamo origin of the G-O rule, it is efficient and physically accurate to reduce the dynamo equations into an iterative map of solar cycle strength \citep{May1976}, as long as the properties of the dynamo are well quantified. \cite{Durney2000,Charbonneau2001,2005ApJ...619..613C,2007ApJ...658..657C} pioneered the use of iterative maps for solar cycle analysis and applied them to study the G-O rule and other properties of solar cycles.  With  recent advancements in solar B-L dynamo research, the iterative map can be revisited and applied to analyze and understand the G-O rule.

In the first article of the series \citep{Wang2025}, we have constructed an iterative map for solar cycles based on observed nonlinearity and stochasticity in the B-L dynamo. The basic component of the iterative map is the mutual generation of poloidal and toroidal fields, generic to solar dynamos.  The specific form of poloidal field generation originates from observation based B-L dynamo nonlinearity and stochasticity in previous works such as \cite{2003SoPh..215...99L, 2008A&A...483..623S, 2010A&A...518A...7D, 2021A&A...653A..27J}. By analyzing the properties of the iterative map and the generated solar cycle series, we have shown that stochasticity is necessarily the primary source of cycle variability for solar dynamo models where the generation of poloidal field from toroidal field follows a growth-and-saturation form. This conclusion holds regardless of specific parameter choices. However, the exact distribution of cycle amplitudes is influenced by the detailed functional form and parameter values of the model. 

In this sequel, we continue to utilize the iterative map to analyze the the G-O rule and its relationship to the nonlinearity and stochasticity of solar dynamo. We give a comprehensive and definitive description of the G-O rule’s various forms and investigate the G-O rule of the generated solar cycles under these forms.  We provide a more general form of the G-O rule, and explain how it is generated from nonlinearity and stochasticity. The results provide implications for theoretical and observational studies on solar dynamo and cycle prediction.

The article is organized as follows. In Section \ref{sec:methods} we review the iterative map that we use to analyze the G-O rule. In Section \ref{sec:results} we examine the quantified results of the G-O rule in varied forms in the iterative map. In Section \ref{sec:explain} we explain the nature of the G-O rule.  We discuss and conclude in Section \ref{sec:outro}.





\section{Reviewing the observation-based iterative map of solar cycles}\label{sec:methods}
\begin{figure}
        \centering
	\includegraphics[width=0.45\textwidth]{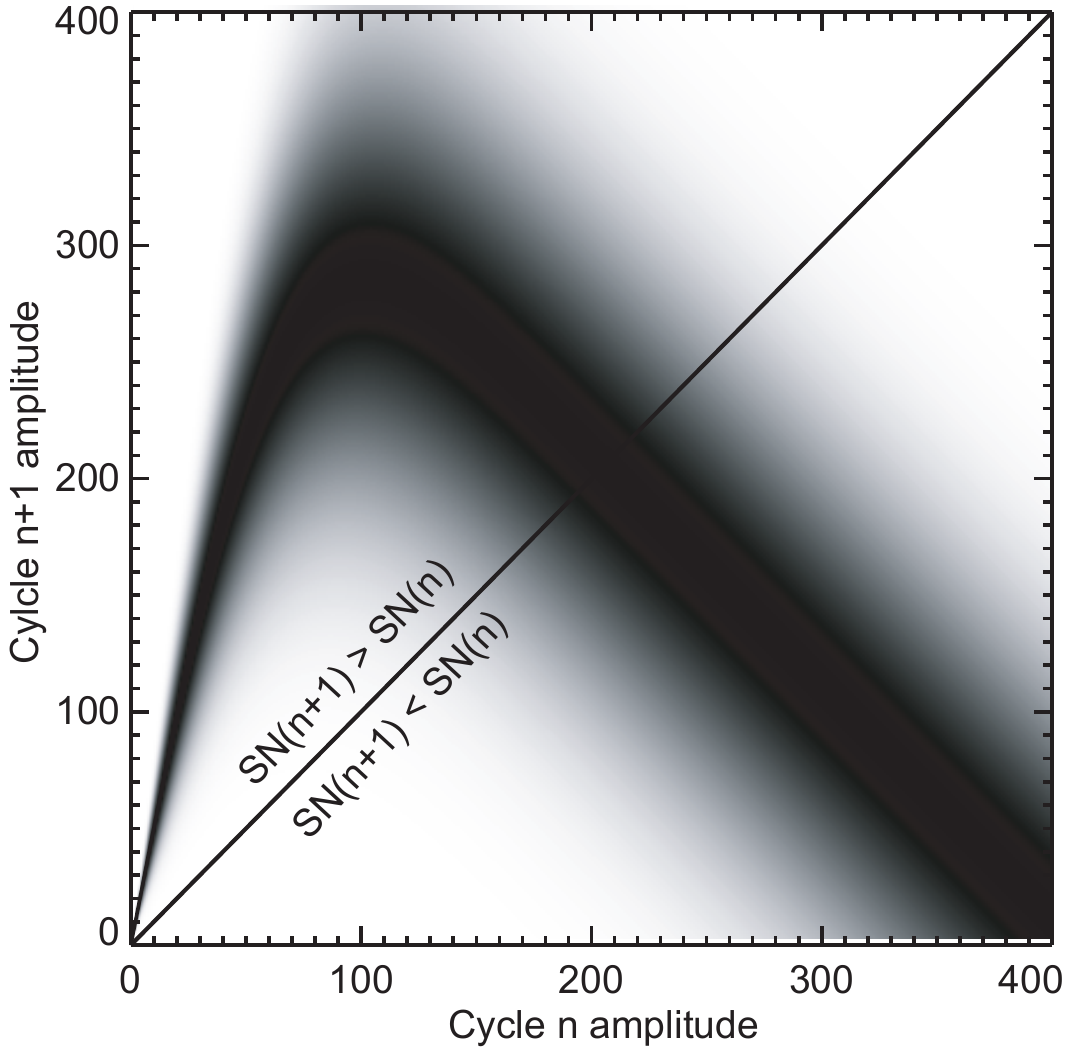}

	\caption{Diagram showing the relationship between cycle amplitude of cycles $n$+1 ($SN (n+1)$) and $n$ ($SN (n)$) as described by Equation (\ref{eq:recursion2}). The gray shaded region represents the distribution of probability density, with darker areas indicating higher likelihood.  The diagonal line divides the plot into two regions: $SN(n+1)>SN(n)$ (upper-left) and  $SN(n+1)<SN(n)$ (bottom-right).\label{fig:rec}}
\end{figure}

The B-L dynamo implies a iterative map of cycle strength, in which the strength of a cycle is determined by the strength of its previous cycle.  We have produced an observational based iterative map in the prequel to this article \citep{Wang2025}, and we review the important points of it briefly here.

\cite{Durney2000} and \cite{Charbonneau2001} first constructed iterative maps of solar cycles by quantifying the mutual generation of poloidal and toroidal field in B-L dynamo.  The poloidal field at the beginning of a cycle can be represented by the strength of polar field as well as the global axial dipole field.  The poloidal field generates the toroidal field by the $\Omega$-effect, which is considered to be mostly linear \citep{1979stp.....2..258O,Schatten1978,2007MNRAS.381.1527J}.  The toroidal field emerges to form active regions, hence the strength of toroidal field can be represented by the strength of solar cycle.  The active regions usually have bipoles tilted against the east-west direction, and contribute net flux to the poles of opposite polarities, which is referred to as the B-L mechanism serving as the means of poloidal field generation from toroidal field.  In the original iterative map of \cite{Durney2000} and \cite{Charbonneau2001}, the newly generated poloidal field from toroidal field directly becomes the poloidal field at the beginning of the next cycle, but as we know that, in the B-L mechanism, the newly generated poloidal field should cancel out the old poloidal field first before building up that at the beginning of the next cycle. This represents one of the key differences between our iterative map and earlier versions.

A significant advancement in understanding the solar cycle over the past decade is the quantification of intrinsic nonlinearity and stochasticity in the B-L mechanism for poloidal field generation, based on direct observations. The tilt angles of active regions during a stronger solar cycle tends to be smaller \citep{2010A&A...518A...7D, 2021A&A...653A..27J}, while the latitudes tends to be higher \citep{2003SoPh..215...99L, 2008A&A...483..623S}.  These effects, referred to as tilt quenching and latitude quenching, respectively, limit the production of poloidal field from toroidal field, and serve as nonlinearity that confines cycle amplitudes in B-L dynamo \citep{Jiang2020, Karak2020, Talafha2022}.  The stochasticity of the B-L dynamo comes from the turbulent convection affecting the rise and emergence of active region field \citep{Weber2013}, resulting in the large scatter of the latitude and tilt of active regions \citep{2011A&A...528A..82J, Jiang2014}.  We use the quantification and parameterization of \cite{Jiang2020}, hereafter J20, to construct the iterative map. The observation-based quantification of  nonlinearity and stochasticity represents another key difference between our iterative map and earlier versions.

With the linear poloidal to toroidal process and the nonlinear and stochastic toroidal to poloidal process, we finally create a recursion function of solar cycles, describing the relationship between the amplitude of cycle $n$+1, denoted as $SN(n+1)$, and the amplitude of cycle $n$, denoted as $SN(n)$.  The recursion function is as follows
\begin{equation}
    SN\left(n+1\right) = k_{0}k_{1}\textbf{erf}\left(\frac{SN\left(n\right)}{quench}\right)\left(1 + stoch\times X\right) - SN\left(n\right),
    \label{eq:recursion2}
\end{equation}
in which $k_{0}$ is the correlation between axial dipole moment at cycle minimum and the next cycle's amplitude, $k_{1}$ and $quench$ are parameters controlling nonlinearity, \textbf{erf} is the error function, $X$ is a normally distributed random variable, and $stoch$ is a parameter setting the standard deviation of stochasticity.  The recursion function is illustrated in Figure \ref{fig:rec}.  While the Gaussian scatter can extend below 0, we limit the values of cycle amplitudes above 0, by employing a reflecting boundary at $SN(n)$=0.  Whenever a negative value of $SN(n)$ occurs, we use its absolute value, reflecting the cycle amplitude back to positive, so that the iterative map can continue.

The recursion function Equation (\ref{eq:recursion2}) has 2 terms on the right-hand side.  The first term is the poloidal field generation from toroidal field.  Considering the property of error function, it first increases as the cycle amplitude $SN(n)$ increases, then saturates after $SN(n)$ is large enough.  The parameter $k_{1}$ is the maximum amount of poloidal field that active regions can generate during a solar cycle in total, while $quench$ controls how fast the poloidal field generation would saturate in terms of $SN(n)$.  The random part $\left(1 + stoch\times X\right)$ is multiplicative to the error function, implying that the actual scatter becomes larger when the poloidal field generation is stronger
-- characteristic of multiplicative noise.  The scatter of poloidal field generation arises inherently from the stochastic nature of active region emergence, with active regions that have large tilt angles and low latitudes contributing most significantly \citep{Jiang2014,2015ApJ...808L..28J, 2018ApJ...863..116W,2020JSWSC..10...46N,2025ApJ...978..147Y}. Its exact form reflects a combination of stochasticity in emergence rate, latitude, area, and tilt of active region emergence. In this study, we adopt the formulation from \cite{Jiang2020}, in which the noise is explicitly multiplicative. While alternative noise formulations may be explored in future studies with more accurate sunspot records, the specific form of the noise does not significantly affect the existence of the G–O rule, as will be discussed in Section \ref{sec:explain}. The first term of Equation (\ref{eq:recursion2}) is subtracted by the second term, indicating that the generated poloidal field should cancel out the old poloidal field.  Hence, the form of the recursion function is a representation of general dynamo processes, and the specific form is determined by observation-based studies.


The parameters of nonlinearity and stochasticity has uncertainty because of observational limitations.  Here, we adopt the parameters of J20 as the standard set of parameters, with polar precursor coefficient $k_0$ being 58.7, maximum dipole moment $k_{1}$ being 6.94, $quench$ being 75.85, and $stoch$ being 0.17, in the following analysis.  We also consider a considerable range of different parameters, evaluating the effect to the G-O rule.  Besides uncertainty of parameters, the form of nonlinearity and stochasticity we adopt from J20 is subject to future refinements as well.

The recursion function, along with an initial cycle amplitude, can be used to produce a large series of solar cycle amplitudes efficiently for the analysis of the G-O rule.  We note that, the aforementioned quantification of nonlinearity and stochasticity is based on cycle amplitude (maximum value of the 13-month smoothed monthly sunspot number over a cycle in Sunspot Number Version 2).   Hence, we use the current recursion function in the following analysis.


\section{Quantification of the G-O rule in different forms in the iterative map}\label{sec:results}
\subsection{Exploring the G-O rule in varied forms using the iterative map}\label{subsec:manifest}
Using the recursion function, i.e. Equation (\ref{eq:recursion2}), we generate a large series of solar cycle amplitudes. We first focus on the cycle strength definition of the G-O rule, so we pair the cycles accordingly. We label the initial cycle as cycle 0, and then pair the cycles sequentially:  0-1, 2-3, 4-5, and so on. Each pair consists of an even cycle with the following odd cycle, which we refer to as G-O pairs. With these pairs, we can evaluate various forms of the quantification of the G-O rule, which are summarized in Table \ref{tab:define}.

We first examine the proportion of pairs where the even cycle amplitude is larger than that of the following odd cycle. This is quantified by calculating the ratio of pairs with a larger even cycle to the total number of pairs, which we refer to as the E-to-A ratio. We generate 1,000,000 cycles, pair them, and calculate the E-to-A ratio, which is found to be $0.4555\pm0.0003$, with the $1\sigma$ uncertainty being the standard error derived from 10 separate E-to-A ratio calculations. This result indicates that there are more pairs with the odd cycle larger than the even cycle. Interestingly, the initial cycle amplitude does not influence the ratio, whether the initial cycle is weak or strong. Furthermore, the starting index for pairing does not affect the outcome: if we label the first cycle as cycle 1, the cycle pairings are reversed, but the result remains the same. Regardless of the pairing method, the latter cycle in each pair consistently has a higher probability of being stronger than the former.

\begin{figure}
	\centering
	\includegraphics[width=0.5\textwidth]{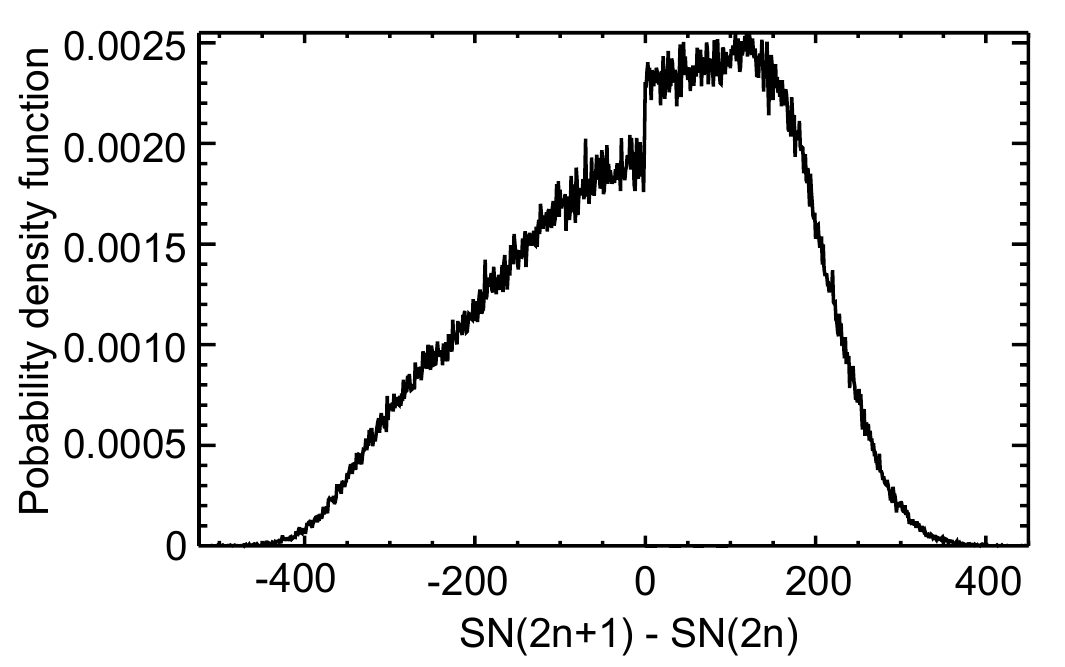}
	\caption{Probability density function (PDF) of the difference between the two solar cycles in each pair, derived from the 1,000,000 cycles generated using the recursion function, i.e., Equation (\ref{eq:recursion2}). The physical origin of the discontinuity at $\Delta SN=0$ is presented in the main text.}
\label{fig:pairs_1}
\end{figure}

Having established that the latter cycles in the G-O pairs tend to be stronger, we now aim to quantify how much stronger they actually are. To do this, we analyze the difference between the two cycles in each pair, denoted as  $\Delta SN = SN(2n+1)-SN(2n)$.  This is similar to the analysis of \cite{2023IAUS..372...70S}.  The probability density function (PDF) of $\Delta SN$ is shown in Figure \ref{fig:pairs_1}, which indicates that $\Delta SN$ follows an asymmetric distribution.  The mean of $\Delta SN$ is nearly 0, while the median is larger than 0, at 19.  The standard deviation is large, at 157.  This indicates that, in the half of the distribution where $\Delta SN>0$, the population is larger, but the values of $\Delta SN$ are smaller. In contrast, for the half where $\Delta SN<0$, the population is smaller, but the values are larger. Therefore, although the latter cycle is more likely to be larger than the former cycle, the expectation of the difference between the cycle strength within the G-O pairs is actually 0. Hence, there is no long-term trend of increasing cycle amplitude.  At exactly $\Delta SN=0$, the distribution is not continuous, but this is not an artifact.  Instead, the reason is that different parts of the recursion function are taken into account when we move $\Delta SN$ from less (below the diagonal line of Figure \ref{fig:rec}) to greater (above the diagonal line of Figure \ref{fig:rec}) than 0.  More details of $\Delta SN$ will be explained in Section \ref{sec:explain}.

\begin{figure}
	\centering
	\includegraphics[width=1\textwidth]{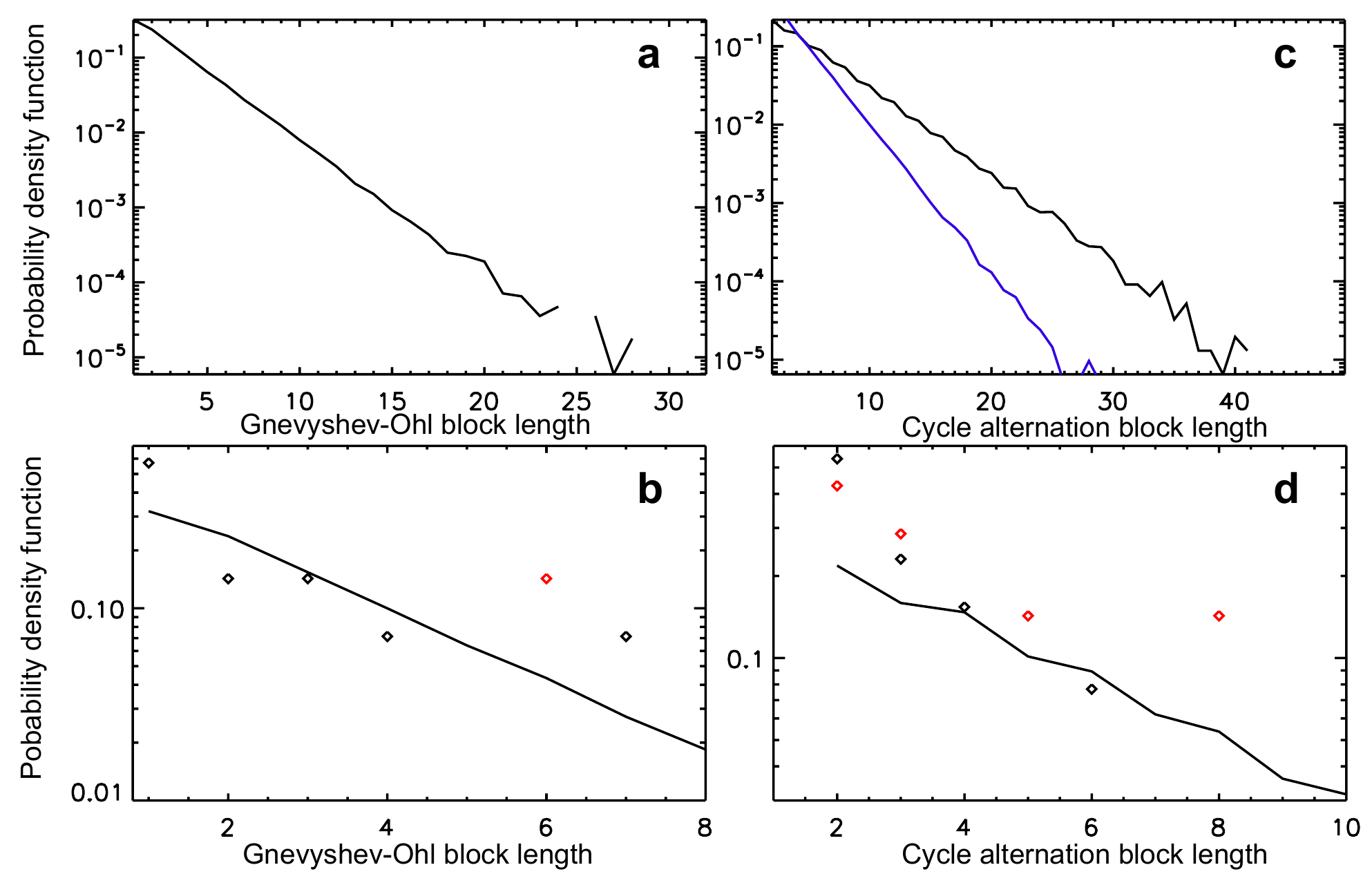}
	\caption{Statistical properties of the G-O pairs and cycle alternation of cycles. (a) The PDF of G-O block length with $y$ axis in logarithmic scale.  (b) A zoom-in of panel (a) in order to compare with the results of \cite{2021A&A...649A.141U} covering cycles from 971 to 1900 (indicated by black diamond symbols), and with the results for cycles 1–24 (indicated by red diamond symbols). The unit of the G-O block length is 1 pair of cycles. (c) The PDF of the cycle alternation block length, with the black curve representing the result of the iterative map, and the blue curve representing the result of a series of fully random cycles.  (d) Same as Panel (b), but for the cycle alternation block length. The unit of the cycle alternation block length is 1 cycle.
\label{fig:pairs_2}}
\end{figure}

The aforementioned values of the E-to-A ratio and the PDF of $\Delta SN$ both have demonstrated the G-O rule holds. This rule likely arises from the nonlinearity and stochasticity inherent in the iterative map, as will be demonstrated in the following two subsections. We then examine the variations in the G-O rule. As reviewed in the introduction, some observations suggest that the G-O rule exhibits long term variations and reversal: some periods follow the G-O rule, while some periods follow a reversed G-O rule. To evaluate this, we define G-O blocks as continuous sequences of G-O pairs in which all pairs have larger even cycles or larger odd cycles, with opposite pairs before and after the block.  We obtain the PDF of the block lengths based on the series of 1,000,000 cycles.  As shown in Figure \ref{fig:pairs_2}a, the length of G-O blocks follows an exponential distribution, which is a natural consequence of the stochastic process.  Theoretically, if an event happens independently at a constant rate, the time interval between two such events will follow an exponential distribution.  In our model, the probability of the pairs switching between even and odd is constant, meaning that the ``reversal'' of the G-O rule is not periodic, but rather random in our model.

Besides the cycle strength definition, the G-O rule is also defined by the correlation definition in observations, which suggests that the E-O correlation is significantly positive, while the O-E correlation is less significantly statistically. However, this is not the case for the 1,000,000 cycles in our model.  Both E-O and O-E correlations are negative, with a value of -0.42, the difference between them being insignificant.  This result is consistent with the fact that the E-to-A ratio does not change if the definition of even or odd is changed.  In fact, the correlations are the same as the correlation between cycle $n$ and $n$+1, in which $n$ is an arbitrary cycle regardless of even or odd.  Since our recursion function only considers the relationship between cycle $n$ and $n$+1, without long term memory, ``even'' and ``odd'' cycles are equivalent in the long term.  However, this does not mean that our model is unrealistic, as the observations of the G-O rule are based on limited number of solar cycles. Before we evaluate and explain the effect of the limited number of solar cycles on the behavior of the G-O rule in detail in Subsection \ref{subsec:lim}, we first investigate the impact of nonlinearity and stochasticity in the following Subsection.

Furthermore, the negative correlations between cycles indicates a tendency toward cycle amplitude alternation. While cycle alternation exhibits different statistical significance from the original formulation of the G-O rule \citep{2024ARep...68...89N}, it is often regarded as closely related.  Therefore, we also analyze it with our model. Unlike \cite{2007ApJ...658..657C} and \cite{2024Univ...10..364P}, who used running means with window widths of 3 or 5 cycles to evaluate cycle alternation, we identify the zigzag pattern directly from the actual cycle amplitudes. This approach is appropriate because our model lacks long-term memory beyond one cycle, as presented in the first paper of this series, making only comparisons between adjacent cycles meaningful. We define a cycle alternation block as a sequence of alternating high and low amplitude cycles that continues until the zigzag pattern is broken. The resulting block lengths are shown in Figure \ref{fig:pairs_2}c. Similar to the G–O blocks, the alternation blocks follow an exponential distribution, reflecting the system’s stochastic nature. For comparison, we also analyze a synthetic series of completely random cycles, similar to the approach of \cite{2007ApJ...658..657C}. These random cycles follow the same PDF of cycle amplitudes from the iterative map, but do not follow any recursion relations. As shown in blue in Figure \ref{fig:pairs_2}c, the alternation blocks in the random series also follow an exponential distribution. However, the iterative map yields significantly more long alternation blocks. This result indicates that cycle alternation is a statistically significant feature of the cycle variability produced by the iterative map.


\begin{table}
    \caption{A summary of varied forms of the G-O rule.}
    \label{tab:define}
    \centering
    \footnotesize
    \setlength{\tabcolsep}{4pt}
    \renewcommand{\arraystretch}{1.2}

   \begin{tblr}{|Q[c,m,0.15\textwidth]|Q[c,m,0.3\textwidth]|Q[c,m,0.3\textwidth]|}
        \hline
        Concept & Definition & Role in the G-O rule \\
        \hline
         Even cycle & cycle with even number (e.g. 0,2,4,...)& The leading cycle in a cycle pair\\
         \hline
         Odd cycle & cycle with odd number (e.g. 1,3,5,...)& The following cycle in a cycle pair\\
         \hline
         G-O pair & a pair consisting of an even cycle and following odd (e.g. 0-1, 2-3, ...)& \\
         \hline
         G-O rule (cycle strength definition) & during a time range, more even cycles are weaker than their following odd cycles & The original form\\
        \hline
         Reversed G-O rule (cycle strength definition) & During a time range, more even cycles are stronger than their following odd cycles & \\
         \hline
         E-to-A ratio & ratio of number of even-odd cycle pairs with larger even cycles to number of all even-odd cycle pairs & A value smaller than 0.5 indicates G-O rule while larger than 0.5 indicates reversed G-O rule\\
         \hline
         G-O block & a series of continuous cycle pairs with larger even (or odd) cycles& An exponential distribution would indicate that the variation of the G-O rule is stochastic\\
         \hline
         $\Delta SN=SN(2n+1)-SN(2n)$ & difference of cycle amplitude within an even-odd cycle pair & The asymmetricity and the median of its distribution imply the G-O rule\\
         \hline
         $\Delta SN=SN(n+1)-SN(n)$ & difference of cycle amplitude between 2 arbitrary adjacent cycles & Similar to above but without cycle pairing, suggesting that cycles are more likely to be followed by a stronger cycle in general \\
         \hline
         E-O correlation & Pearson's correlation between even cycles and following odd cycles& The correlation definition \\
         \hline
         O-E correlation & Pearson's correlation between odd cycles and following even cycles& The correlation definition suggests the E-O correlation tends to be larger than the O-E correlation for a certain time range\\
        \hline
         Cycle alternation & a tendency of cycle amplitudes following an alternative weak-strong pattern & It is often considered highly related to the original definition of the G-O rule\\
        \hline
    \end{tblr}
\end{table}

\subsection{Impact of Nonlinearity and Stochasticity on the Behavior of the G-O Rule}\label{subsec:params}
The key components of the iterative map are the B-L nonlinearity and stochasticity, so they should play a certain role in the G-O rule.  In order to know how nonlinearity and stochasticity affects the G-O rule, we use different parameters, and observe how the E-to-A ratio and $\Delta SN$ of the G-O rule are affected.  This is also meaningful due to the fact that the parameters themselves are of uncertainty.  Based on the standard set of parameters, we consider a $\pm 25\%$ variation of $k_{1}$, $quench$, and $stoch$, changing one parameter at a time while keeping others unchanged. In total, 6 cases are considered: $0.75\times k_{1}$, $1.25\times k_{1}$, $0.75\times quench$, $1.25\times quench$, $0.75\times stoch$, and $1.25\times stoch$.  We also include an optimized set of parameters from our prequel paper, which is $k_{1} = 6.94\times0.9, quench = 75.85\times1.5$, and $stoch = 0.17\times0.75$.  This set of parameter produces a PDF of cycle amplitude, which is closer to the normal cycles component in \cite{2014A&A...562L..10U} compared to the standard set, but it is not guaranteed to be the best fit to observations.

We first calculate different E-to-A ratios for the aforementioned 6 parameter sets, shown in Table \ref{tab:goratio}.  The uncertainty of each E-to-A ratio is obtained from the standard error of 10 individual solar cycle series generated by the method.  All produce a ratio below 0.5, confirming the validity of the G-O rule.  Meanwhile, the E-to-A ratio reacts to parameter changes differently.  Larger maximum dipole moment $k_{1}$ makes the ratio closer to 0.5, so the two types of pairs would be more similar, which means that the G-O rule is weaker.  Larger $quench$ or $stoch$ decreases the ratio, enlarging the difference between the two types of pairs, indicating a stronger presence of the G-O rule.  The optimized set also produces a smaller E-to-A ratio compared to the standard set, which indicates that the G-O rule might be stronger in observations.

We also calculate the different $\Delta SN$ values.  We have explained in Subsection \ref{subsec:manifest} that the mean of $\Delta SN$ is 0 for the reason that the cycle strength needs to be confined, so here we show the medians of $\Delta SN$, as we did above.  As shown in Table \ref{tab:goratio}, all cases produce medians larger than 0, with varied values.  While larger median of $\Delta SN$ can be regarded as stronger G-O rule, this is not necessarily the case, as larger $k_1$ produces weaker G-O rule from the perspective of E-to-A ratio, but produces larger median of $\Delta SN$.  This is because the median of $\Delta SN$ is not only related to how many $\Delta SN$ values are larger or smaller than 0, but also related to the shape of the distribution.  Different forms of the G-O rule may act differently when the parameters change.

Usually, the G-O rule, particularly its manifestation as cycle alternation, is explained by nonlinearity \citep[e.g.,][]{2007ApJ...658..657C}, which produces a strong cycle after a weak cycle and vice versa, thus forming a semi-regular pattern, while stochasticity is often considered to be destructive to semi-regular behavior.  Our analysis actually shows that while the form of nonlinearity affects the G-O rule, the stochasticity is constructive to the G-O rule as well, under the original definition of the G-O rule. The form and parameters of nonlinearity and stochasticity are both important to the G-O rule.  

\begin{table}
    \caption{Impact of the parameters in nonlinearity and stochasticity on the behavior of the G-O rule based on the iterative map.}
    \label{tab:goratio}
    \centering
    \footnotesize
    \setlength{\tabcolsep}{4pt}
    \renewcommand{\arraystretch}{1.2}
    \begin{tabular}{lccc}
        \hline
        Parameter set & E-to-A ratio & Median of $\Delta$SN\\
        \hline
         Standard set & $0.4555\pm0.0003$ &$19.1 \pm 0.1$\\
         $0.75\times k_{1}$ & $0.4371\pm0.0002$&$16.26\pm 0.07$\\
         $1.25\times k_{1}$ & $0.4653\pm0.0002$&$20.97\pm 0.09$\\
         $0.75\times quench$ & $0.4672\pm0.0005$&$16.4\pm0.1$\\
         $1.25\times quench$ & $0.4414\pm0.0002$&$21.49\pm0.08$\\
         $0.75\times stoch$  & $0.4754\pm0.0002$&$10.7\pm0.1$\\
         $1.25\times stoch$ & $0.4425\pm0.0003$&$24.8\pm0.1$\\
         Optimized set &  $0.4492\pm0.0002$&$11.23\pm0.06$\\
        \hline
    \end{tabular}
\end{table}

\subsection{Impact of a Limited Number of Solar Cycles on the Behavior of the G-O Rule}\label{subsec:lim}
\begin{figure}
        \centering
        \includegraphics[width=1\textwidth]{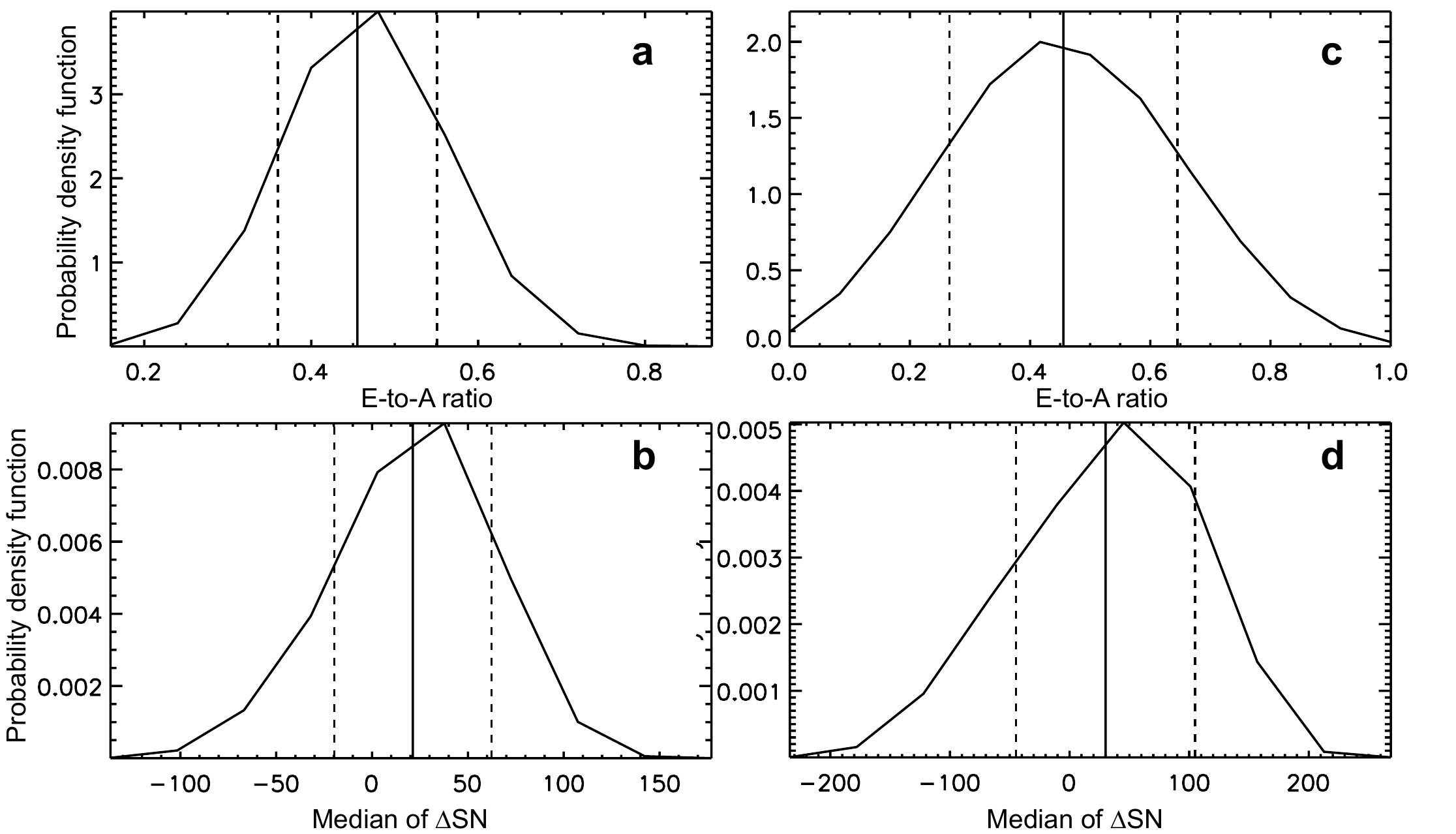}
	\caption{Statistical properties of the G-O rule under limited numbers of cycles. Panels (a) and (b) are the PDF of the E-to-A ratio and the median of $\Delta SN$, respectively, for 50 pair of cycles. The solid and the dashed vertical lines represent the mean and $1\sigma$ range, respectively. Panels (c) and (d) are same as (a) and (b), but for the number of cycle pairs being 12. }
    \label{fig:limited}
\end{figure}

\begin{figure}
                \centering
        \includegraphics[width=1\textwidth]{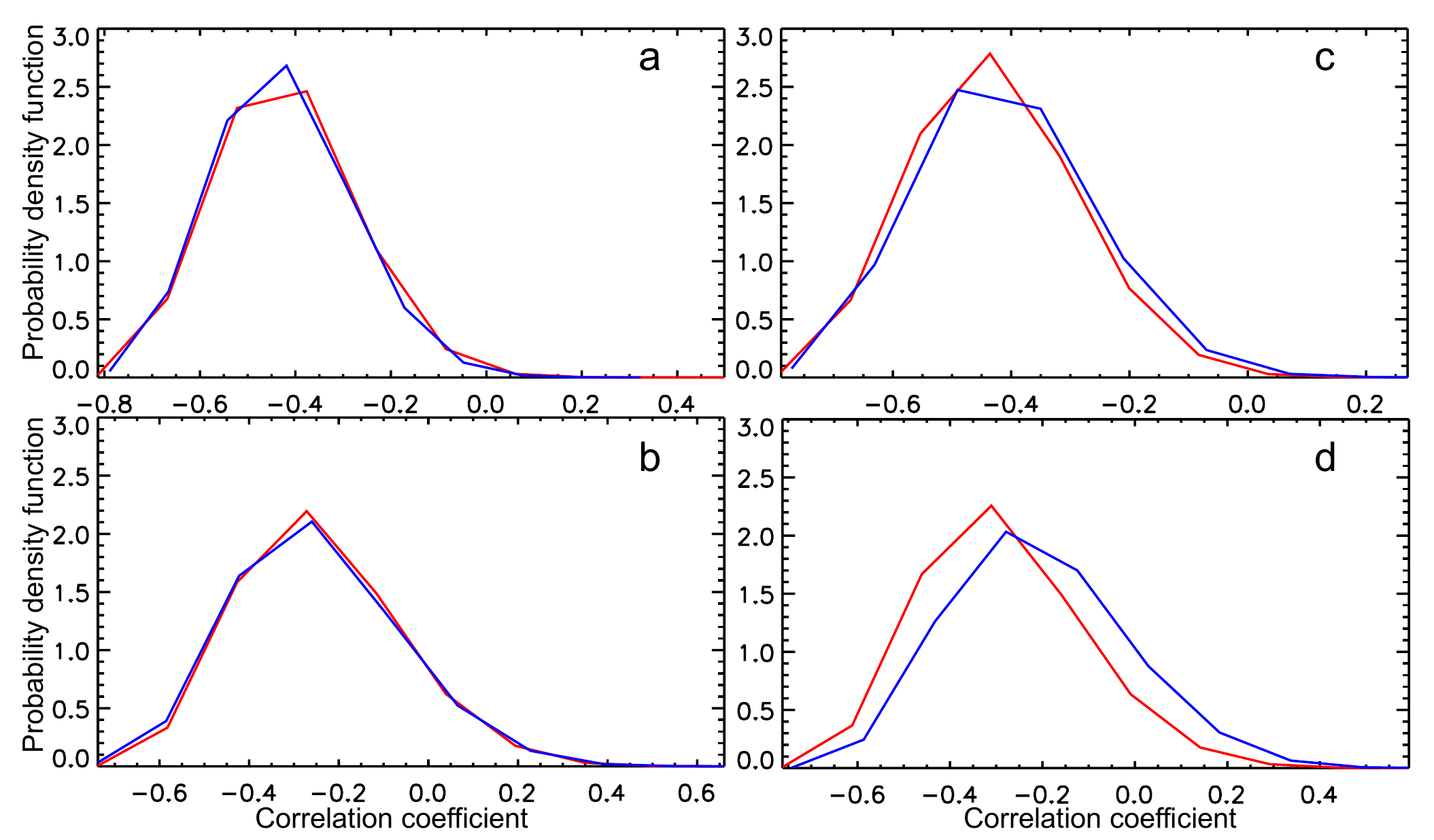}
	\caption{Statistical properties and examples of the correlation definition of the G-O rule for 50 pairs of even and odd cycles. Panels (a) and (b) are the PDF of correlation coefficients under the standard set of parameters and the optimized set of parameters, respectively, with the initial cycle amplitude 81.2. The red curve indicates the E-O correlation, while the blue curve indicates the O-E correlation.  Panels (c) and (d) are same as panels (a) and (b), respectively, but for the initial cycle amplitude 285. }
    \label{fig:corrdef}
\end{figure}

\begin{figure}
                \centering
        \includegraphics[width=1\textwidth]{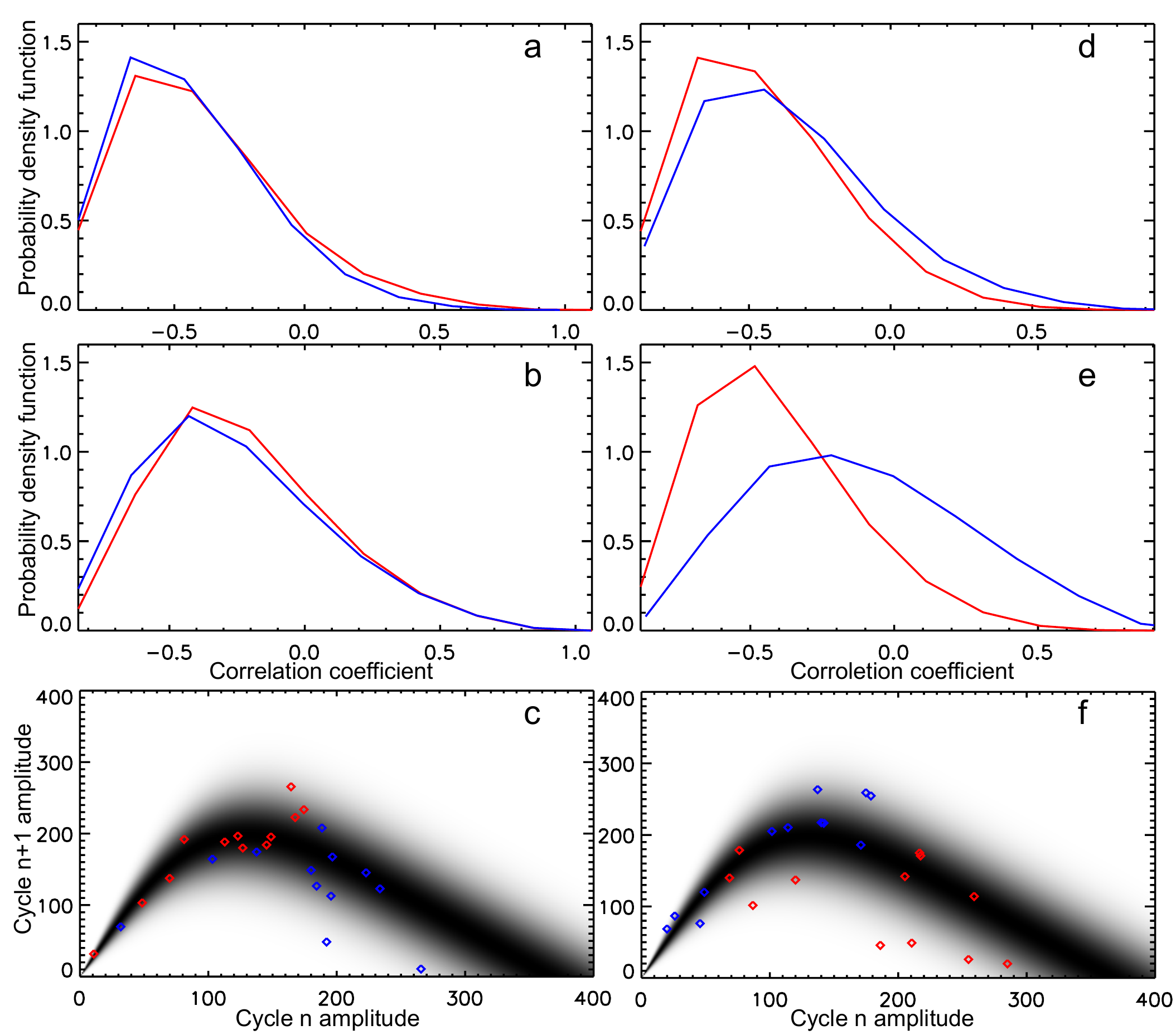}
	\caption{Statistical properties and examples of the correlation definition of the G-O rule for 12 pairs of even and odd cycles. Panels (a)-(c) correspond to the case with initial cycle amplitude 81.2. Panels (a) and (b) are the PDF of correlation coefficients under the standard set of parameters and the optimized set of parameters, respectively. The red curve indicates the E-O correlation, while the blue curve indicates the O-E correlation.  (c) A set of 24 cycles whose E-O correlation is larger than 0.9 is taken as an example for the relationship between cycle $n$ and cycle $n$+1 amplitudes. The red (blue) symbols indicate $n$ being even (odd). Panels (d)-(f) are same as panels (a)-(c), respectively, but for the initial cycle amplitude 285.}
    \label{fig:corrdef2}
\end{figure}

The aforementioned results on the G-O rule are valid when the number of cycles is sufficiently large. However, real-world observations are based on a limited number of well-resolved solar cycles, which necessitates understanding how the G-O rule behaves with fewer cycles. We consider two time ranges for solar cycles.  The first spans 100 cycle, i.e., 50 G-O pairs, comparable to the millennial analysis of \cite{2023IAUS..372...70S} based on radioisotopes. The second covers 24 cycles, corresponding to the era of directly observed sunspot records.  By analyzing a large set of such cycle series, we aim to derive the properties of the G-O rule when the number of cycles is limited.



We first examine whether there are more pairs in which the former cycle is larger than the latter cycle. We generate 100,000 sequences, each consisting of 100 cycles, and compute the E-to-A ratio for each sequence. The resulting 100,000 E-to-A ratios are used to construct the PDF shown in Figure \ref{fig:limited}a. Although the mean of the PDF is slightly below 0.5, it lies within $1\sigma$ of 0.5. This indicates a weak tendency for the cycles to follow the G-O rule, but not at a statistically significant level. When we decrease the total number of cycles to 24, the distribution becomes broader, while the mean value remains unaffected, as shown in Figure \ref{fig:limited}c. This helps explain the inconsistent manifestation of the G-O rule across different segments of observed solar cycles. Based on ISN version 2.0, for solar cycles 2-25, corresponding to a random realization of 24-cycle series, its E-to-A ratio is 0.25, which lies slightly outside the $1\sigma$ range of the 100,000 realizations. 

We also examine $\Delta SN$ under limited cycle numbers.  As stated in Section \ref{subsec:manifest}, the median of $\Delta SN$ is larger than 0, indicating that more cycles are followed by stronger, rather than weaker, ones. Hence, here we produce the PDF of the median of $\Delta SN$, shown in Figure \ref{fig:limited}b for the time range of 100 cycles.  Again, the median of $\Delta SN$ is larger than 0, but still not larger than the $1\sigma$ range, which again implies that the G-O rule is not statistically significant. This also supports the result of \cite{2023IAUS..372...70S}, who argue that for the reconstructed solar cycles, the invalidity of the Gnevyshev-Ohl rule cannot be regarded as an evidence of its insufficient accuracy. It could reflect the inherently weak statistical significance of the G-O rule over the considered time period. Similar to the E-to-A ratio, the distribution of the median of $\Delta SN$ also broadens when we limit the cycle number to 24, as shown in Figure \ref{fig:limited}d. For cycles 2–25, the median $\Delta SN$ is 45.7 based on ISN version 2.0, a value that remains well within the expected distribution. Compared to the behavior of the E-to-A ratio, this result further explains the conflicting manifestations of the G–O rule reported in earlier studies, which often relied on different formulations of the rule.

Now, we compare the results of the G-O block with those from observed limited cycles. We compare the PDF of the G-O block shown in Figure \ref{fig:pairs_2}a with the results obtained from pairing cycles between 971 and 1900 reconstructed by \cite{2021A&A...649A.141U} shown in Figure \ref{fig:pairs_2}b.  We choose all cycles from their Table 1, excluding the grand minima cycles, and pair up the rest, and obtain the length of each G-O block.  We note that this is a brief comparison, and the varied data qualities in \cite{2021A&A...649A.141U} are not considered.  We also include the blocks during cycle 1-25, marked in red.  Overall, the PDF follows the reconstructed cycle data.  However,  the relatively small number of observed solar cycles and their uncertainty limit the ability of observational data to effectively constrain our model, making this a preliminary comparison.

Similarly, we compare the results of cycle alternation with those from observed limited cycles.  Again, the results from \cite{2021A&A...649A.141U} and the results of cycles 1-25 are used.  As shown in Figure \ref{fig:pairs_2}d, both are roughly comparable to the results of the iterative map, but the probabilities are in general larger than the iterative map. The total number of observed cycle alternation blocks is small. In contrast, the iterative map generates 1,000,000 cycles, which even includes blocks longer than 10, albeit rarely. The deviation between the observed and the results from the iterative map likely arises from the limited number of observed cycles, representing only a short segment of a random realization.

We continue to examine the correlation definition of the G-O rule with a limited number of solar cycles.  As we have shown, in the long run, the relationship between adjacent cycles will certainly follow Equation (\ref{eq:recursion2}), regardless of even or odd, so we expect that the correlation definition is only applicable when cycle numbers are limited. We speculate that the initial amplitude of the limited cycle series is important.  To test this, we let all 100,000 sets of 100 cycles to start at two amplitudes, $SN(0) = 81.2$, and $SN(0) = 285$, which are the weakest and strongest cycles since Cycle 1, correspondingly. Then, we observe the PDF of the E-O correlation and the O-E correlation. The PDFs of these correlations with $SN(0) = 81.2$ are shown in Figures \ref{fig:corrdef} (a) and (b) for the iterative map using the standard and optimized parameter sets, respectively. The difference between the two correlations are not notable. For the case with $SN(0) = 285$, shown in Figures \ref{fig:corrdef} (c) and (d), the difference between the two correlations are more significant for the optimized set of parameters. All these plots indicate that both correlations are not quite likely to be significantly positive.  Instead, there is a higher probability for negative values. 

We further restrict the analysis to 24 cycles, with the results shown in Figure \ref{fig:corrdef2}. The difference between low (left panels) and high (right panels) initial cycle amplitudes becomes more pronounced. For an initial amplitude of 81.2, the positive correlations, consistent with observations and presented in Section \ref{subsec:manifest}, are located in the tail of the PDFs. The corresponding E-O correlation is slightly larger than the O-E correlation, again consistent with observation. To understand the observed correlation definition of the G-O rule, we show a set of 24 cycles as an example in Figure \ref{fig:corrdef2}(c), in which the E-O correlation is significantly positive, larger than 0.9. In this example, most even-odd pairs (red) fall on the rising part of the recursion function, while odd-even pairs (blue) do not. Sine the rising part of the recursion function is relatively linear, correlations are higher when the pairs of cycles lie in this range. In contrast, when the initial amplitude is 285, the O-E correlation tends to be stronger, and the difference between E–O and O–E correlations becomes more evident for the optimized set of parameters.  Figure \ref{fig:corrdef2}(f) shows another example where the O–E correlation exceeds 0.9. Here, most odd–even pairs (blue) lie within the linear region of the recursion function. These results suggest that the correlation definition is sensitive to the initial cycle amplitude.

The higher O-E correlation observed over limited time range presented above arises from the asymmetric shape of the recursion function between the rising and descending parts. The rising phase tends to be more linear with less scatter, while the descending phase is more nonlinear and exhibits greater variability. As a result, one correlation is more strongly influenced by cycle pairs along the linear (rising) part of the recursion function, whereas the other is dominated by the more scattered, nonlinear (descending) part. When the initial cycle in a pair has an extreme amplitude, the difference between the E-O and O-E correlation coefficients becomes more pronounced.

In summary, for limited cycles, the G-O rule by its cycle strength definition is only a trend and is not statistically significant even at $1\sigma$ significance.  The behavior of solar cycles are more likely to form different G-O blocks, with some blocks having larger odd cycles, some having larger even cycles, and the total E-to-A ratio is not guaranteed to be either larger or smaller than 0.5.  This favors the observational studies suggesting that there are variations of G-O rule, such as \cite{2001SoPh..198...51M,2013ApJ...772L..30T,2015Ge&Ae..55..902Z}.  We note that the variations of G-O rule in our iterative map is random instead of systematic.  As for the correlation definition of the G-O rule, it is in general a little more likely for E-O correlation to be larger than O-E correlation when the initial cycle amplitude is low, but still the correlation definition is not guaranteed under a limited number of cycles.


\section{The general form of the G-O rule and its origin from the nonlinearity and stochasticity}\label{sec:explain}

\begin{figure}
        \centering
        \includegraphics[width=0.5\textwidth]{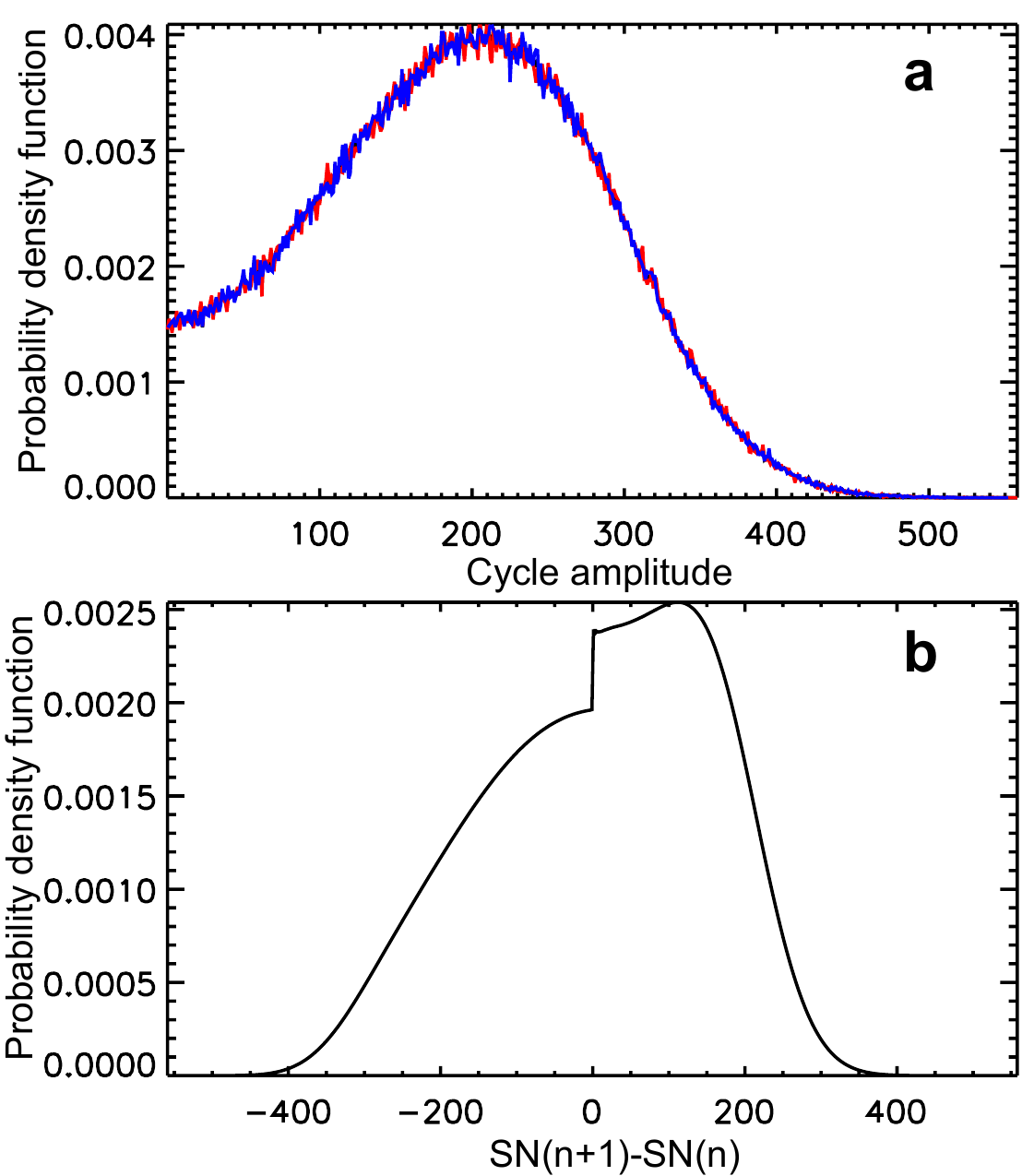}
	\caption{Probability density function of cycle amplitude (upper panel) and $\Delta SN$ (lower panel). The curve in the upper panel is actually composed of three overlaid curves, which are a black curve referring to all cycle amplitude, a red curve indicating even cycle amplitude, and a blue curve showing odd cycle amplitude. For the physical origin of the discontinuity at $\Delta SN=0$, see the corresponding main text.}
    \label{fig:int}
\end{figure}

In the previous section we have analyzed the different quantified forms of the G-O rule, which enable us to investigate the intrinsic origin of the G-O rule in this section.  From Subsection \ref{subsec:manifest}, we know that the E-to-A ratio is unaffected by the method of pairing cycles.  This implies that there is an inherent property of the cycles that does not require pairing them at all.  In fact, the iterative map, which describes the relationship between cycle $n$ and $n$+1, does not distinguish between even and odd cycles from the outset.  As discussed in the first paper of the series \citep{Wang2025}, any solar cycle in the iterative map loses all its initial information after just a few iterations and becomes indistinguishable from all cycles in the statistical sense.  This means that even and odd-numbered cycles are not distinguishable.  This is true, as Figure \ref{fig:int}a shows that even and odd-numbered cycles have the same PDF.

From this perspective, the general G-O rule in the long run does not need pairing at all.  For any two adjacent cycles, there is a larger probability for the latter cycle to be stronger than the former cycle, which is referred to as the general form of the G-O rule.  We evaluate this in the following.  Let $p(x)$ be the PDF of $x$ with $x$ being $SN\left(n\right)$, and $q(y)$ be the PDF of $y$ with $y$ being $SN\left(n+1\right)$. The former $p(x)$ is an unconditional probability, and the latter $q(y)$ is a conditional probability $P(y|x)$, which is a measure of the probability of $SN\left(n+1\right)$ occurring, given the strength of $SN\left(n\right)$. Then the general G-O rule is to examine the following probability
\begin{equation}
    P\left(y>x\right) = \int_{x=0}^{\infty}\int_{y=x}^{\infty}q\left(y\right)p\left(x\right)dydx.
\end{equation}
The function $p(x)$ should be obtained by generating a large number of cycle amplitudes with Equation (\ref{eq:recursion2}), and the result is presented in Figure \ref{fig:int}(a).  The function $q(y)$ as a conditional probability, is not the same as $p(x)$, but is determined by the recursion function, as the nonlinearity determines its peak and the stochasticity determines its scatter.  In our case, the stochasticity is a Gaussian distribution, so $q(y)$ should be as follows,
\begin{equation}
    q\left(y\right)=\frac{1}{\sqrt{2\pi}\sigma_{y}}\exp\left(-\frac{\left(y-<y>\right)^{2}}{2\sigma_{y}^{2}}\right),
\end{equation}
where $<y>=k_{0}k_{1}\textbf{erf}\left(\frac{x}{quench}\right)-x$, and $\sigma_{y}=k_{0}k_{1}\textbf{erf}\left(\frac{x}{quench}\right)\times stoch$. This probability can be calculated as long as we have the PDF $p(x)$. Using the PDF in Figure \ref{fig:int}(a), which is from Section 3.2 of \cite{Wang2025}, we have $P\left(y>x\right)=0.546$.  Then, the E-to-A ratio is 0.454, very close to the E-to-A ratio in Subsection \ref{subsec:manifest}.

Now we do not consider $\Delta SN$ within pairs of cycles, but $\Delta SN$ of two arbitrary adjacent cycles.  The PDF of $\Delta SN$,  $P\left(y=x+\Delta SN\right)$, is actually as follows
\begin{equation}\label{eq:int}
    P\left(y=x+\Delta SN\right) = \int_{x=0}^{\infty}q\left(x+\Delta SN\right)p\left(x\right)dx .
\end{equation}
The result of the integration of Equation (\ref{eq:int}) is presented in Figure \ref{fig:int}b, which is identical to Figure \ref{fig:pairs_1}, showing an asymmetric distribution on two sides and a discontinuity at 0.  The discontinuity at 0 is a result of the property of the recursion function shown in Figure \ref{fig:rec}.  The integration Equation (\ref{eq:int}) can be regarded as an integration of the recursion function along a line $y = x + \Delta SN$, weighted by the PDF of cycle amplitude.  As it is shown in  Figure \ref{fig:rec}, when $\Delta SN$ is smaller than 0, such a line only covers the descending part of the recursion function.  But when $\Delta SN$ becomes larger than 0, the ascending part of the recursion function is also included. The additional ascending component leads to a discontinuous and asymmetric distribution. In particular, the portion near the origin results in the noticeable jump at $\Delta SN = 0$.

Surely enough, the long term G-O rule of the iterative map is a direct result of the nonlinearity and stochasticity in the B-L mechanism.  The recursion function in Figure \ref{fig:rec} is different in the $\Delta SN$ less or greater than 0 part, then the long-term G-O rule will occur.  Such difference make it more likely for a weak cycle to be followed by a stronger cycle, but less likely for a strong cycle to be followed by a weaker cycle.  From this perspective, not limited to the recursion function we introduce, Other nonlinear stochastic iterative maps could also produce the G-O rule. But the exact statistical significance varies according to the specific form and parameters. When the nonlinearity and stochasticity change, how much of the recursion function falls to the upper-left part and the lower-right part in Figure \ref{fig:rec} becomes different, and the significance of G-O rule is different as well.  We note that, ``cycles are more likely to be followed by a stronger cycle'' does not result in unbounded growth of cycle strength.  The cycles in this statement is an arbitrary cycle among the PDF of cycles.  For a definite cycle, whether the next cycle is probably stronger or weaker is solely determined by Equation (\ref{eq:recursion2}).  Statistically, while $\Delta SN$ is more likely to be positive, the absolute values of positive $\Delta SN$ are smaller than negative, so the cycle amplitude is not unbounded.

\section{Discussion and conclusions}\label{sec:outro}
In this article, we have analyzed the G-O rule using an observation-based iterative map developed in the prequel to this study \citep{Wang2025}. A larger portion of the recursion function lies in the region where the cycle amplitude increases from one cycle to the next, rather than decreases. When this recursion function is weighted by the PDF of cycle amplitudes and integrated over the full amplitude range, it implies that an arbitrary solar cycle is statistically more likely to be followed by a stronger cycle than by a weaker one. This underlying asymmetry represents a generalized form of the G-O rule, unifying the various forms reported in the literature when solar cycles are analyzed in pairs. Over sufficiently long timescales, explicit cycle pairing becomes unnecessary to observe this trend. On shorter periods lasting a millennium or less, the G-O rule manifests only as a weak trend without statistical significance. During such intervals, the solar cycle can randomly alternate between following the G-O rule and its reversed form, without a strong preference for either, consistent with observational studies that report temporal variations in the rule. The exact tendency of the G-O behavior under a limited number of cycles is influenced by the form and parameters of the recursion function, which should be taken into account in future investigations of solar cycle nonlinearity and stochastic dynamics.

The observation-based iterative map for solar cycles introduced in the prequel \citep{Wang2025}, though seemingly simple, is an effective tool for investigating the generic and complex behaviors of nonlinear systems, like the solar cycle. The recursion function incorporates fundamental solar dynamo processes: the regeneration between toroidal field and poloidal field, and the cancellation between opposing poloidal polarities. We have clearly shown that the G-O rule in its general form originates from the nonlinearity and stochasticity of poloidal field generation, generic to solar dynamos.  The quantification of the G-O rule, especially under limited solar cycle number,  is heavily affected by the form and parameterization of the nonlinearity and stochasticity, hence more realistic observations are key to evaluate the observed G-O rule.

In the iterative map that generates the G-O rule, one cycle solely determines the next, and there is no longer-than-1-cycle memory.  From this perspective, it is not necessary to consider the Hale 22-year cycle as the fundamental component of solar cycle evolution in order to explain the G-O rule, nor is other long-term memory or fossil field needed.  Still, we do not completely rule out the possibility of other explanations.

Since we interpret the G-O rule as an integration of the recursion function weighted by the PDF of cycle amplitudes, we can show that the G-O rule is a direct product of nonlinearity and stochasticity.  Because of nonlinearity, the recursion function is intrinsically different in its $SN(n+1)>SN(n)$ and $SN(n+1)<SN(n)$ part, with stochasticity included.  Then, there will be a statistical difference in the positive and negative part of the distribution of $\Delta SN$.  Such a result is not limited to the specific nonlinearity and stochasticity we introduce, yet the exact statistical significance is affected by the specific formations. The G-O rule in the long term is a natural result of nonlinearity affected by stochasticity.

The statistical properties can be vastly different when the number of cycles is limited.  Observational studies, such as \cite{2015SoPh..290.1851H}, tend to show that the correlation coefficient between adjacent cycles is positive.  The correlation definition of the G-O rule suggests that the correlation is more significantly positive for even-odd pairs than odd-even pairs.  We have shown that such coefficients can have large scatter in their distribution when the number of cycles is 100, and is largely dependent on parameters of nonlinearity and stochasticity, and dependent on initial cycle amplitude.  If the starting cycle is weak, and the weak part of the recursion function is close to linear function, then it is possible for the correlation definition of the G-O rule to occur in a limited range.  If the number of cycles is as limited as the directly observed cycles, the uncertainty of the G-O rule will become even larger, making it less meaningful for direct comparison.  Better observation-based nonlinearity and stochasticity should provide more accurate explanation for characteristics of the observed short-term G-O rule.


Our iterative map at present does not include grand minima, as it has been discussed in the Discussions section in our first paper \citep{Wang2025}.  Proxies of long-term solar activity show that grand minima refer to a separate peak in the probability density function along with the normal cycles \citep{2014A&A...562L..10U}.  The original definition of the G-O rule \citep{1948G_O}, starting from cycle 1, does not include grand minima; more recent observational evaluations like \cite{2023IAUS..372...70S} do not consider grand minima either.  From the perspective of evaluating and understanding the G-O rule, reconstructing realistic grand minima is not obligatory.  On the other hand, while large temporal ranges of high solar activity, referred to as grand maxima exist, their exact difference to normal cycles and physical origin are not clear, for they do not appear to have a distinctive peak in observational probability density function \citep{2023LRSP...20....2U}.  From this perspective, we do not explicitly regard strong cycles as grand maxima in our iterative map when trying to understand the G-O rule.

Another extreme case associated with grand minima and maxima is the occurrence of large deviations from the expectation value in some random realizations.  In our model, we adopt a reflecting boundary to keep the cycle amplitudes positive.  If not, some negative values will occur due to the large randomness, in which case the total amount of dipole moment generated during the cycle is too small to cancel out the dipole from the previous cycle and build up the dipole field of opposite polarity at the start of the next cycle.  The stochasticity in our model mainly originates from the randomness in the latitude and tilt of active regions. As established by \cite{Jiang2014}, active regions emerging at low latitudes with large tilts can have a profound effect on cycle variability. In extreme cases, rogue active regions with large flux, low latitude, and anti-Joy's tilt can halt the dynamo process by preventing the build up of the poloidal field of the next cycle, as shown in the simulation examples of \cite{2017SoPh..292..167Nagy}. This physical scenario corresponds to the possibility of negative values in the iterative map.  Better understanding of such extreme cases and better treatments in the recursion function will be considered in future studies.


Future advance in the solar cycle recursion relation within the framework of the B-L dynamo will be able to provide better explanation for the G-O rule. Such progress will depend on both more accurate observations of solar cycles and a deeper understanding of the underlying B-L dynamo mechanisms. In the foreseeable future, the number of well-resolved solar cycles is unlikely to be sufficient for determining the E-to-A ratio with statistical significance. As a result, the G-O rule should be viewed as a statistical trend rather than a strict law within the B-L dynamo framework. Consequently, it should not be used in isolation as a definitive observational constraint in solar dynamo modeling, nor as a standalone predictive tool for future solar cycles. A more reliable approach to understanding and applying the G-O rule lies in combining it with other characteristics of solar cycle evolution.

While the number of cycles is too limited for the E-to-A ratio, several observational studies of the G-O rule can be conducted to examine the nonlinearity and stochasticity based G-O rule.  The first is the revision of the observational $\Delta$SN \citep[e.g.,][]{2023IAUS..372...70S}.  As we have already pointed out in the manuscript, its median instead of mean value is more fitted to represent the G-O rule.  Another valuable topic is the distribution of the G-O blocks and cycle alternation blocks, from better and longer time scale data.  Whether these blocks follow exponential distribution is a direct examination of the stochastic nature of the G-O rule.  Furthermore, if the evaluation of the correlation definition can be extended to $\sim$50 pairs of cycles in the future with enough statistical significance, it will be more likely to examine whether the currently observed difference in correlation coefficients is an effect of limited cycle numbers or not.

\begin{acknowledgements}

The international sunspot number version 2.0 is from the World Data Center SILSO, Royal Observatory of Belgium, Brussels, which is available at https://www.sidc.be/SILSO/datafiles. This research was supported by the National Natural Science Foundation of China through grant Nos. 12425305, 12350004, 12173005, 12373111, \& 12273061, the National Key R\&D Program of China through grant No. 2022YFF0503800, and supported by Specialized Research Fund for State Key Laboratory of Solar Activity and Space Weather.
\end{acknowledgements}

\bibliography{ref}{}
\bibliographystyle{raa}
\end{document}